\documentclass[aps,prl,nofootinbib,twocolumn,superscriptaddress,preprintnumbers]{revtex4}

\usepackage{amssymb}
\usepackage{amsmath}
\usepackage{epsfig}
\usepackage{hyperref}
\usepackage{breakurl}

\makeatletter
\def\simgt{\mathrel{\lower2.5pt\vbox{\lineskip=0pt\baselineskip=0pt
           \hbox{$>$}\hbox{$\sim$}}}}
\def\simlt{\mathrel{\lower2.5pt\vbox{\lineskip=0pt\baselineskip=0pt
           \hbox{$<$}\hbox{$\sim$}}}}
\makeatother

\def\mysection#1{{\bf #1.} }

\newcommand{\be}{\begin{equation}}
\newcommand{\ee}{\end{equation}}
\newcommand{\bea}{\begin{eqnarray}}
\newcommand{\eea}{\end{eqnarray}}
\newcommand{\beq}{\begin{eqnarray}}
\newcommand{\eeq}{\end{eqnarray}}

\def\mysection#1{{\bf #1.} }

\def\lsim{\mathrel{\rlap{\lower4pt\hbox{\hskip1pt$\sim$}}
     \raise1pt\hbox{$<$}}}         %less than or approx. symbol
\def\gsim{\mathrel{\rlap{\lower4pt\hbox{\hskip1pt$\sim$}}
     \raise1pt\hbox{$>$}}}         %greater than or approx. symbol

\begin{document}

\title{Cosmology and time dependent parameters induced by misaligned light scalar}

\author{Yue Zhao}
\affiliation{Tsung-Dao Lee Institute, and Department of Physics and
Astronomy,\\ Shanghai Jiao Tong University, Shanghai, 200240,
China.} \affiliation{Michigan Center for Theoretical Physics,
University of Michigan, Ann Arbor, MI 48109}

\begin{abstract}
We consider a scenario where time dependence on physical parameters
is introduced by the misalignment of an ultra-light scalar field.
The initial VEV of such field at the early time remains a constant
until Hubble becomes comparable to its mass. Interesting
cosmological consequences are considered. Light sterile neutrinos
hinted by terrestrial neutrino experiments are studied as a
benchmark model. We show the BBN constraints can be easily avoided
in this scenario, even if reheating temperature is high. The scalar
can be naturally light in spite of its couplings to other fields.
Parameters of sterile neutrino may remain changing with time
nowadays. This can further relax the tension from the recent IceCube
constraints.
\end{abstract}

\maketitle

%%%%%%%%%%%%%
\mysection{Introduction} Measurements from cosmology may provide
important information or impose strong constraints on possible
extensions to the Standard Model (SM). For example, dark matter
thermal relic abundance may be used to extract information in the
dark sector. Alternatively, if the dark sector contains light
particles which have sizable couplings to SM sector, it could be
disfavored due to measurements like $N_{eff}$ \cite{Ade:2015xua}.

On the other hand, a non-trivial evolution of the dark sector during
the history of the Universe is able to introduce time dependence to
physics parameters, which indicates that conclusions from
cosmological measurements may not be applied, in a straightforward
manner, to physics measured in our local solar system today.

In this letter, we consider a theory with an ultra-light scalar
field $\phi$. The VEV of $\phi$ is assumed to be related to the
masses of certain fields, e.g. a fermion $\psi$. We assume $\phi$
gets a VEV at the beginning of the Universe
\cite{Preskill:1982cy,Abbott:1982af,Dine:1982ah}. When Hubble is
larger than $m_\phi$, the field value remains approximately
unchanged, which will be referred as ``early time'' in later
discussion. The field starts oscillating and its VEV decreases when
Hubble becomes smaller than $m_\phi$. The time dependence of VEV
could have interesting cosmological implications. \footnote{Other
models where certain parameters have non-trivial time dependence
have been considered in
\cite{Dimopoulos:1990ai,Anderson:1997un,Fardon:2003eh,Cohen:2008nb}.
The change of VEV is either introduced by a phase transition or a
chemical potential by populating particles coupled to the scalar
field. }

This scenario can be applied generically. For example, if the dark
matter mass and/or interaction change with time, it may be
non-trivial to interpret the calculation of thermal relic abundance
to what it implies in our local experiments, such as DM (in)direct
detections. This has been considered in the content of O(keV) or
heavier sterile neutrino DM \cite{Berlin:2016bdv} \footnote{Their
focus is on heavy sterile neutrino with mass O(keV) or higher. The
mixing with active neutrino is large when the Universe if hot in
order to produce proper relic abundance, and it becomes small to
avoid indirect detections nowadays, such as X/$\gamma$-ray line
searches.}. The other possibility is for light dark matter particles
having sizable couplings to SM particles. They may be disfavored
from cosmological points of view, such as $N_{eff}$ measurements.
However, a time dependent mass and interaction of DM induced by
$\phi$'s evolution can easily relax the tensions. Thus such light DM
should not be dismissed by simply implementing the cosmological
arguments \cite{Hochberg:2015pha,Hochberg:2015fth}. At last, if the
oscillation of the light scalar field in our solar system still
plays a role on changing physical parameters, it can introduce time
dependence into the experimental results. Searching for that induced
by an oscillating dilaton field as DM background, has been studied
in
\cite{Arvanitaki:2014faa,VanTilburg:2015oza,Arvanitaki:2015iga,Arvanitaki:2016fyj}.
Related to neutrino properties, \cite{Berlin:2016woy} considers the
scenario where the scalar field VEV introduces additional mixing
among active neutrinos. This can be constrained by the null results
from anomalous periodicities measurements in the solar neutrino
flux.

In this letter, we are focused on sterile neutrinos with masses
O(eV) and mixing angles to active neutrinos at O(0.1). These choices
are motivated by the anomalies reported in short distance $\bar
\nu_\mu\to\bar\nu_e$ flavor conversion measurement at the LSND
experiment
\cite{Athanassopoulos:1996jb,Athanassopoulos:1996wc,Louis:1995vg},
as well as other terrestrial neutrino experiments such as MiniBooNE
\cite{Aguilar-Arevalo:2012fmn}. The preferred parameters of sterile
neutrinos are in strong tension with cosmological measurements such
as nucleosynthesis and large scale structure \cite{Giunti:2015wnd}.
Many efforts have been devoted to reconcile these tensions, for
example, by late entropy production \cite{Gelmini:2004ah},
additional interactions to sterile neutrino \cite{Cherry:2016jol},
non-trivial neutrino number density dependence in the mass matrix
\cite{Ghalsasi:2016pcj}, or late time phase transition in the dark
sector \cite{Chacko:2004cz,Davoudiasl:2017jke}. A more comprehensive
review can found in \cite{Abazajian:2012ys,Adhikari:2016bei}.

In our setup, we introduce a light scalar field $\phi$ which obtains
a VEV as initial condition. We further introduce a coupling between
$\phi$ and sterile neutrino $\psi$, so that the VEV of $\phi$ has
non-trivial contribution to the Majorana mass term of $\psi$. In the
current local solar system, $\phi$'s VEV is much smaller than that
during BBN. Thus the mass and mixing of sterile neutrino obtain
strong time dependence. We will demonstrate that the constraint from
BBN can be efficiently relaxed in this setup.

\mysection{Introducing $\phi$ dependence to Fermion mass} We
consider the following coupling between a light scalar $\phi$ and a
sterile neutrino $\psi$
\begin{eqnarray}\label{eq:LangPsi}
L\supset (m_0+g'^2\frac{\phi^2}{M})\psi\psi.
\end{eqnarray}
This particular coupling can be easily realized in a UV model. For
example, $\phi$ may carry a $Z_2$ parity and its coupling to $\psi$
is induced by integrating out some heavy scalar.

In principle, this model can be further simplified if we do not
include the mass term $m_0$ or do not impose the $Z_2$ symmetry of
$\phi$. However, we make these choices in order to avoid the
subtlety that $\psi$ becomes much lighter than $\phi$ when
$\langle\phi\rangle$ becomes small during oscillation. Such
phenomenon is studied as parametric resonance production and it is
considered in
\cite{Traschen:1990sw,Kofman:1994rk,Shtanov:1994ce,Kofman:2004yc}.
Consequently, energy density in $\phi$ are lost through the
production of $\psi$, which will further back react to the evolution
of $\phi$. Though this additional subtlety may have important
consequences if it happens, this deviates the main focus of this
letter.

\mysection{Cosmological evolution of a light scalar field} Depending
on the detailed history of $\phi$, it may or may not have a non-zero
initial field value away from its minimum before inflation. For
simplicity, let us assume the initial field value in the patch of
our current Universe before inflation is a universal constant
$\phi_{init}$.

During inflation the field value will be perturbed away from its
universal initial value by quantum fluctuations. The power spectral
density is
\begin{eqnarray}
P_\phi(k)=\sigma_\phi^2=\bigg(\frac{H_{inf}}{2\pi}\bigg)^2.
\end{eqnarray}
Thus the generic value of $\phi$ randomly fluctuates between
$\phi_{inf}\in(\phi_{init}-\frac{H_{inf}}{2\pi},
\phi_{init}+\frac{H_{inf}}{2\pi})$.

If $\phi$ does not have strong interactions with other fields, its
field value remains as a constant after inflation, until Hubble
becomes comparable to the mass, i.e. $H_{osc}\simeq m_\phi$. After
the oscillation starts, $\phi$ behaves as matter, and its energy
density scales as $a^{-3}$, where $a$ is the scale factor of the
Universe.

This light scalar field may or may not play the role of DM. If
$\phi$ composes O(1) fraction of DM, there is a lower bound on its
mass, i.e. $10^{-22}$ eV
\cite{Turner:1983he,Press:1989id,Sin:1992bg,Hui:2016ltb}. This gives
a lower bound on the temperature of the Universe at which the scalar
field starts oscillating,
\begin{eqnarray}\label{eq:DMosc}
H_{osc}|_{\textrm{min}}\simeq\frac{\textrm{keV}^2}{M_{pl}}\simeq
10^{-22} \textrm{eV}.
\end{eqnarray}
This is still before matter-radiation equality. Thus the energy
density is properly parameterized by the temperature of the
Universe, and $H_{osc}$ can be written as
\begin{eqnarray}\label{eq:Tosc}
T_{osc}\sim \sqrt{m_\phi M_{pl}}.
\end{eqnarray}

On the other hand, if $\phi$ is not the dominant contribution to DM,
its mass can be even lower, and it could start oscillating at a
later time.

The average energy density of DM as a function of time can be
written as
\begin{eqnarray}
\bar\rho_{DM}(t)\simeq 10^{-6} \frac{1}{a(t)^3}
\textrm{GeV}/\textrm{cm}^3 \simeq 0.6
\bigg(\frac{T(t)}{\textrm{eV}}\bigg)^3\textrm{eV}^4,
\end{eqnarray}
Here we take the average DM energy density in the current Universe
to be $10^{-6} \textrm{GeV}/\textrm{cm}^3$. In the last equation, we
use the approximation that the temperature of the Universe scales as
an inverse linear function of the scale factor, neglecting possible
 modifications from entropy dumping. To have a consistent
cosmology, we require the energy density in $\phi$ when it starts
oscillating to be the same or smaller than that of DM during that
time. More explicitly, we have
\begin{eqnarray}
\frac{1}{2}m_\phi^2 \phi_{inf}^2 \le 0.6
\bigg(\frac{T_{osc}}{\textrm{eV}}\bigg)^3\textrm{eV}^4.
\end{eqnarray}
If $\phi$ starts oscillating during radiation dominated era, Eq.
(\ref{eq:Tosc}) is applicable and we get
\begin{eqnarray}\label{eq:phiInf}
\phi_{inf}&\le& 10^{18}
\bigg(\frac{10^{-22}\textrm{eV}}{m_\phi}\bigg)^{1/4} \textrm{GeV}.
\end{eqnarray}

Here we emphasize that the calculation above is assuming $\phi$
evolves as a free field. One may be worried that the existence of
particles coupling to $\phi$, such as $\psi$, may contribute as an
effective chemical potential of $\phi$, thus modifies its evolution
when the Universe cools down, e.g. in
\cite{Anderson:1997un,Ghalsasi:2016pcj}. However, if the production
of $\psi$ is suppressed by either a large mass or small coupling
induced by the large VEV of $\phi$ at the early time, then it has
little impact on the evolution of $\phi$ and it is consistent to
treat $\phi$ as a free field.

\mysection{$N_{eff}$ during BBN}\label{sec:Neff} Let us consider a
scenario where $\psi$ has sizable interaction with SM particles. If
the properties of this fermion remain the same during the history of
the Universe, it can be thermally populated. If its mass during BBN
is smaller than MeV, it contributes to $\Delta N_{eff}$, $m_0$ and
its coupling to SM would be strongly constrained.

One typical scenario is light sterile neutrino with large mixing
angle. The existence of such sterile neutrino may explain the
long-standing experimental anomaly in short distance
$\bar\nu_\mu\to\bar\nu_e$ flavor conversion
\cite{Athanassopoulos:1996jb,Athanassopoulos:1996wc,Louis:1995vg,Aguilar-Arevalo:2012fmn}.
The experimental results cannot be properly explained if there are
only three neutrino flavors. For a recent summary, please see
\cite{Gonzalez-Garcia:2015qrr}. On the other hand, if sterile
neutrinos are added, the measurements favor a squared mass
splitting,  i.e. $\Delta m^2$, around O(1) $\textrm{eV}^2$ and a
mixing angle with active neutrino as $\theta\sim $ O(0.1).
\footnote{In the following discussion, we only consider one active
neutrino and one sterile neutrino. The generalization to multiple
species is straightforward.} The equilibrium of sterile neutrinos
with the SM thermal bath can be reached as long as the reheating
temperature is only slightly higher than the electron-active
neutrino decoupling temperature, i.e. around 1 MeV. This is in
tension with the measurements \cite{Ade:2015xua} which gives
$N_{eff}=3.15\pm0.23$.

On the other hand, if there is a non-trivial dependence on
$\langle\phi\rangle$ for the sterile neutrino mass, its mass at the
early Universe can be very different from its current value, which
matters for terrestrial neutrino experiments. Thus the constraints
from the thermal history of the Universe, e.g. $N_{eff}$, may not be
trivially applied.

First, we calculate the local value of $\langle\phi\rangle$ in our
solar system. If the de Broglie wavelength of $\phi$ is smaller than
the scale of our galaxy, it behaves as a particle from structure
formation point of view. We expect the ratio between local energy
density of $\phi$ to its current average value in our Universe to be
the same as that of DM, i.e.
\begin{eqnarray}
\frac{\rho_{\phi,\odot}}{\bar\rho_{\phi,0}}\simeq\frac{0.3\
\textrm{GeV}/\textrm{cm}^3}{10^{-6}\
\textrm{GeV}/\textrm{cm}^3}\simeq 10^5.
\end{eqnarray}
Thus we can estimate the local VEV of $\phi$ as
\begin{eqnarray}
\rho_{\phi,\odot}&\simeq& 10^5\times
\frac{1}{2}m_\phi^2\phi_{inf}^2\bigg(\frac{T_0}{T_{osc}}\bigg)^3\nonumber\\
&\simeq& 10^{-6}\times
m_\phi^2\phi_{inf}^2\bigg(\frac{\textrm{eV}}{T_{osc}}\bigg)^3,
\end{eqnarray}
which indicates
\begin{eqnarray}
\frac{\phi_\odot}{\phi_{inf}}\simeq 10^{-3}\times
\bigg(\frac{\textrm{eV}}{T_{osc}}\bigg)^{3/2}.
\end{eqnarray}
For example, if $m_\phi\sim 10^{-22}$ eV, $T_{osc}$ is about keV.
This indicates that the VEV of $\phi$ during the early Universe can
be about 8 order to magnitude larger than that locally in our solar
system.

%\begin{eqnarray}
%\sigma_{\phi,\odot} \simeq 10^{10} \bigg(\frac{10^{-22}
%\textrm{eV}}{m_\phi}\bigg)\textrm{GeV}
%\end{eqnarray}

In order to obtain some intuition, let us consider some benchmark
numbers. First, we would like $\psi$ to obtain a mass larger than at
least $10$ MeV in order not to be produced in thermal bath if the
reheating temperature barely triggers BBN\footnote{Later we will see
that this is not necessary for sterile neutrino since the the mixing
is also largely suppressed when $\langle\phi\rangle$ is large.},
i.e.
\begin{eqnarray}\label{eq:Mvalue}
(m_0+g'^2\frac{\phi_{inf}^2}{M})\simeq
g'^2\frac{\phi_{inf}^2}{M}>10\ \textrm{MeV}.
\end{eqnarray}
Here we assume $m_0$ is positive and much smaller than
$g'\frac{\phi_{inf}^2}{M}$ or else the change on $\phi$'s VEV during
the evolution of the Universe cannot make a difference.

Now let us consider two limits, $m_0\ll
g'^2\frac{\phi_{\odot}^2}{M}$ and $m_0\gg
g'^2\frac{\phi_{\odot}^2}{M}$. When $m_0\ll
g'^2\frac{\phi_{\odot}^2}{M}$, the current fermion mass in our solar
system is related to that in the early Universe as
\begin{eqnarray}
\frac{m_{\psi,\odot}}{m_{\psi,inf}}=\frac{\langle\phi_{\odot}\rangle^2}{\langle\phi_{inf}\rangle^2}\simeq
 10^{-6}\times
\bigg(\frac{\textrm{eV}}{T_{osc}}\bigg)^{3}.
\end{eqnarray}
On the other hand, if $m_0\gg g'^2\frac{\phi_{\odot}^2}{M}$, the
contribution from the VEV of $\phi$ in our solar system is
negligible, which simply implies $m_{\psi}$ being larger than that
in the scenario where $m_0\ll g'^2\frac{\phi_{\odot}^2}{M}$. Thus in
summary, we have
\begin{eqnarray}
\frac{m_{\psi,\odot}}{m_{\psi,inf}}\ge 10^{-6}\times
\bigg(\frac{\textrm{eV}}{T_{osc}}\bigg)^{3}.
\end{eqnarray}

For sterile neutrino, we need the local $\psi$ mass to be O(1) eV.
Then $m_{\psi}$, during early time of the Universe, can be easily
larger than $10$ MeV. More explicitly, if $m_{\phi}\sim 10^{-22}$ eV
and $m_{\psi,\odot}$ is about 1 eV, $m_{\psi,inf}$ can be as large
as PeV.

\mysection{Coupling as a function of $\langle\phi\rangle$} So far,
we only consider how $\langle\phi\rangle$ affects $m_\psi$. At the
meanwhile, it also affects the mixing between sterile and active
neutrinos. Let us consider a simple supersymmetric theory,
\begin{eqnarray}\label{eq:natural4}
W\supset \frac{1}{2}m_0 \Psi^2+y H L \Psi + \frac{g'}{2
M}\Phi^2\Psi^2+\frac{1}{2}m_\phi \Phi^2.
\end{eqnarray}
Here $\Phi$ is the supermultiplet containing $\phi$, $\Psi$ contains
$\psi$ and its superpartner $\tilde\psi$. $H$ and $L$ are the higgs
and lepton supermultiplets in MSSM.

The mixing angle can be written as $\theta\sim y v/ m_\psi$. To fit
the anomalies in terrestrial neutrino experiments, we have
$m_{\psi,\odot}\sim$ eV and $y\sim 10^{-12}$. However, during the
early Universe, $m_\psi$ is much larger than its current value in
our solar system, which implies a much smaller mixing angle.

Let us estimate how the suppression on the mixing angle may change
the production of sterile neutrinos. The weak interaction (WI)
collision rate is $\Gamma_{WI}\sim n\sigma\sim G_F^2 T^2 T^3$, while
the oscillation rate goes as $\Delta m^2/T$. One can determine the
cross over point as
\begin{eqnarray}
T_{cross}\sim (\Delta m^2/G_F^2)^{1/6}.
\end{eqnarray}
When temperature is higher than $T_{cross}$, Quantum Zeno effect is
important \cite{Stodolsky:1986dx,Cherry:2016jol} and the flavor
conversion rate can be written as
\begin{eqnarray}
P(\nu_a \to \nu_s)\sim \textrm{sin}^2
2\theta\times\bigg(\frac{\Delta m^2}{T\ \Gamma_{WI}}\bigg)^2.
\end{eqnarray}
Comparing to Hubble expansion rate, in order to be in equilibrium,
one needs
\begin{eqnarray}
T_{high}\le (\textrm{sin}^2 2\theta \frac{\Delta
m^4}{G_F^2}M_{pl})^{1/9}.
\end{eqnarray}
Here $T_{high}$ indicates the temperature at which the thermal
equilibrium can be reached assuming it is higher than $T_{cross}$.
When the temperature is lower than $T_{cross}$, Quantum Zeno effect
is not important. The averaged conversion probability can be written
as
\begin{eqnarray}
\bar P(\nu_a \to \nu_s)=\frac{1}{2} \textrm{sin}^2 2\theta.
\end{eqnarray}
Then the equilibrium can be reached when
\begin{eqnarray}
T_{low}\ge (G_F^2 M_{pl} \textrm{sin}^2 2\theta)^{-1/3}.
\end{eqnarray}
In order to avoid the constraints from $N_{eff}$, we need
$T_{low}>T_{high}$. This indicates
\begin{eqnarray}
(\theta^2\ \Delta m\ G_F\ M_{pl})<1.
\end{eqnarray}
Taking the approximation that $\Delta m\sim m_\psi$ and $\theta\sim
\frac{0.1 \textrm{eV}}{m_\psi}$, we get
\begin{eqnarray}
m_\psi> \textrm{keV}.
\end{eqnarray}

In summary, one may resolve the tension between $N_{eff}$ and the
preferred parameters of sterile neutrino in two ways. One is to
simply raise the sterile neutrino mass to be higher than reheating
temperature. One can also suppress the sterile neutrino production
rate by reducing its mixing angle to active neutrinos. It turns out
that the second choice is more effective. If $m_\psi$ is heavier
than keV before/during BBN, sterile neutrinos are not thermally
populated even with a high reheating temperature.

\mysection{Naturalness of $\phi$'s mass} $\phi$ being ultra-light is
crucial in our scenario. However $\phi$ has non-trivial coupling to
$\psi$. Thus one needs to check whether it is natural to expect
$\phi$ to have such small mass.

The 1-loop contributions are quadratically divergent
\begin{eqnarray}\label{eq:natural1}
\delta m_\phi^2\sim
\frac{g'^2}{16\pi^2}\frac{\langle\phi\rangle^2}{M^2}(\Lambda^2-m_\psi^2).
\end{eqnarray}
Here we truncate the quadratic divergences at a scale $\Lambda$ and
assume $m_\psi$ is dominated by $\phi$'s VEV. $\Lambda$ is supposed
to be the scale where additional physics comes in and cancel the
quadratic divergences from $\psi$'s loop. One typical example is to
identify $\Lambda$ as the mass of superpartner of $\psi$. Before the
oscillation of $\phi$, we have $m_\psi\simeq
g'\langle\phi_{inf}\rangle^2/M$. Thus by requiring naturalness of
$m_\phi$, Eq. (\ref{eq:natural1}) implies
\begin{eqnarray}\label{eq:natural2}
(m_{\tilde\psi}^2-m_\psi^2)&\le&16\pi^2\frac{m_\phi^2}{m_\psi^2}
\langle\phi_{inf}\rangle^2.
\end{eqnarray}
We require $m_\psi$ to be at least keV before the oscillation of
$\phi$. If $\phi$ starts oscillating during the radiation dominated
era, one can use Eq. (\ref{eq:phiInf}) to estimate the misalignment
of $\phi$. This gives
\begin{eqnarray}\label{eq:natural3}
(m_{\tilde\psi}^2-m_\psi^2) &\le&
\bigg(\frac{m_\phi}{10^{-22}\textrm{eV}}\bigg)^{3/2}
\bigg(\frac{\textrm{keV}}{m_\psi}\bigg)^{2}(\textrm{keV})^2.
\end{eqnarray}

In order to avoid the constraints from BBN, we need $\phi$ to start
oscillating at temperature below O(MeV). This gives an upper limit
to $\phi$'s mass, i.e. $m_\phi\sim 10^{-16}$ eV. Plugging into Eq.
(\ref{eq:natural3}) and taking $m_\psi$ to be keV, naturalness
requires
\begin{eqnarray}\label{eq:natural6}
(m_{\tilde\psi}^2-m_\psi^2)\sim (10\ \textrm{MeV})^2.
\end{eqnarray}
Such degeneracy implies a very small SUSY breaking effects in the
dark sector. However this is not impossible to achieve since the
dark sector is mostly isolated from SM sector.

Take the superpotential in Eq. (\ref{eq:natural4}), for simplicity,
let us assume the current sterile neutrino mass in our solar system
is dominated by $m_0$, i.e. $m_0\sim$ eV. To achieve a mixing angle
of O(0.1), we need $y\sim 10^{-12}$. Such tiny coupling between
$\Psi$ and SM supermultiplets introduces a SUSY breaking mass to
$\Psi$ as
\begin{eqnarray}\label{eq:natural5}
(m_{\tilde\psi}^2-m_\psi^2)\sim\frac{y^2}{16\pi^2}(100\
\textrm{GeV})^2\sim (10^{-2} \textrm{eV})^2.
\end{eqnarray}

Another possible contribution to the mass splitting between
$\tilde\psi$ and $\psi$ is from the non-zero VEV of $\phi$. Since
$\phi$ is not strictly flat, its displacement away from the origin
could contribute a positive vacuum energy and break SUSY. It is
straightforward to show that the mass difference between $\psi$ and
$\tilde\psi$ due to the non-zero VEV of $\phi$ can be written as
\begin{eqnarray}
(m_{\tilde\psi}^2-m_\psi^2)\sim m_\psi m_\phi,
\end{eqnarray}
which is much smaller than that in Eq. (\ref{eq:natural6}).

At last, the gravity mediated SUSY breaking effects are unavoidable.
A low scale SUSY breaking is phenomenologically allowed in scenario
of gauge mediation, where $F\sim$ O(10) TeV$^2$. Such SUSY breaking
effects may further introduce a mass splitting between $\tilde\psi$
and $\psi$ at O(meV), which indicates
\begin{eqnarray}
(m_{\tilde\psi}^2-m_\psi^2)\sim m_\psi\ \textrm{meV}\sim
(\textrm{eV})^2,
\end{eqnarray}
where $m_\psi$ is taken to be keV at the last step.\footnote{One may
be worried about the gravity mediated SUSY breaking effects directly
apply to $\phi$. However, $\phi$ may have an approximate shift
symmetry, which is only broken by its small mass term and
interaction with $\psi$. Similar argument has been used in, for
example, relaxion models \cite{Graham:2015cka}.}

In summary, SUSY may be introduced to stabilize $\phi$'s mass, and
the superpartner of $\psi$ cannot be too heavy. Such a requirement
is not impossible since $\psi$ couples very weakly to the rest of
the theory.  We emphasize that the naturalness is not a necessary
criteria to satisfy, rather a subjective requirement.

\mysection{Comparing with ``late time neutrino mass'' models} The
idea on time-dependent sterile/active neutrino mass matrix is not
new. Similar ideas have been explored in the ``late time neutrino
mass'' models \cite{Chacko:2004cz}. Such models consider a
possibility that a phase transition happens after BBN and generates
both active and sterile neutrinos' masses at late time. Such phase
transition is introduced by additional light scalar fields, and it
could be triggered by the decrease of thermal masses of the scalar
fields. Thus the temperature in dark sector can neither be zero nor
equal to that in SM sector, but a little bit lower.

The phase transition spontaneously breaks global symmetries and
light/massless goldstone bosons appear in low energy spectrum. In
order to fit the anomalies in terrestrial neutrino experiments, the
couplings among active neutrino, sterile neutrino and the goldstone
modes are not negligible. Thus the active neutrinos will recouple
with the dark sector when temperature is low, i.e. $T_{rec}\sim$
O(eV). Furthermore, the mean free paths of neutrinos may also be
modified due to their additional interactions with goldstone bosons.

These complications do not happen in our scenario. Since our light
scalar field $\phi$ does not directly talk to active neutrinos,
there is no process can induce recoupling between active neutrino
and dark sector. The change of field's VEV happens automatically
after Hubble becomes smaller than its mass. Thus we do not need the
potential of our scalar field to change with time, and the
temperature in the dark sector can be simply zero.

\mysection{Conclusion} In this letter, we study a model to relax the
cosmological constraints on O(eV) sterile neutrino with O(0.1)
mixing with active neutrino, by introducing a late oscillating light
scalar field.

$\phi$ may be still oscillating in our solar system. If its effects
remain important nowadays, e.g. when $m_0\ll
g'\langle\phi_\odot\rangle^2/M$, physical parameters may still
change with time. If $m_\phi$ ranges from $10^{-22}$ eV to
$10^{-16}$ eV, the period is about seconds to years. This introduces
non-trivial time-dependence into experimental results.

Amusingly, the recent result from IceCube \cite{TheIceCube:2016oqi}
disfavors sterile neutrino parameters from global fits
\cite{Conrad:2012qt,Kopp:2013vaa}. However, it is important to note
that IceCube data is mainly in tension with LSND, but remains
consistent with MiniBooNE. Given the fact that the operating time of
IceCube partially overlaps with that of MiniBooNE but very different
from that of LSND, introducing time dependence may resolve this
tension. A detailed analysis to include time dependence in global
fits could be interesting and we leave it for future study.

%%%%%%%%%%%%%%%%%%%
\mysection{Acknowledgments}
The ideas on introducing time-dependence to physical parameters in
order to relax BBN constraints for light DM were formed in common
with Yonit Hochberg and Kathryn Zurek. Initial trials on models were
developed together. I am grateful to Bibhushan Shakya for useful
discussions, especially Aaron Pierce for readings of the manuscript.
YZ is supported by DOE grant DE- SC0007859.

\end{document}